\documentclass[twocolumn,english]{article}
\usepackage[T1]{fontenc}
\usepackage[latin9]{inputenc}
\usepackage[letterpaper]{geometry}
\usepackage{graphicx}
\usepackage{esint}
\usepackage{babel}

\begin{document}

\title{Individual and Collective Behavior of Small Vibrating Motors Interacting
Through a Resonant Plate}

\author{David Mertens \and Richard Weaver}
\maketitle
\begin{abstract}
We report on experiments of many small motors -- cell phone vibrators
-- glued to and interacting through a resonant plate. We find that
individual motors interacting with the plate demonstrate hysteresis
in their steady-state frequency due to interactions with plate resonances.
For multiple motors running simultaneously, the degree of synchronization
between motors increases when the motors' frequencies are near a resonance
of the plate, and the frequency at which the motors synchronize shows
a history dependence.
\end{abstract}

\section{Introduction}

Ensembles of oscillators that spontaneously synchronize have been
studied for decades. Biological examples abound\cite{BuckQRevBiol1938,BushComplexity2006,WinfreeJTheorBiol1967},
but synchronization occurs in many other systems, including coupled
metronomes\cite{PantaleoneAmJPhys2002,UlrichsChaos2009}, laser arrays\cite{KourtchatovPhysRevA1995},
chemical oscillators\cite{KissScience2002}, arrays of convective
cells \cite{MirandaIntJBifurcationChaos2010}, Josephson junctions
arrays \cite{WeisenfeldPhysRevE1998}, transport networks\cite{KincaidComplexity2008},
and perhaps most notoriously, pedestrians crossing the Millennium
Bridge in London when it first opened\cite{StrogatzNature2005}. These
systems are all examples of populations of similar but not identical
oscillators that exhibit the same basic patterns of behavior, that
(1) they synchronize spontaneously, without the need for any external
driving, and (2) as the oscillators' coupling increases, their synchronization
strengthens. For an overview of the topic, see the review by Acebron
et al. \cite{AcebronRevModPhys2005} and the popular book \emph{Sync},
by Strogatz\cite{Strogatz_Sync}.

The topic of synchronization is much broader than the study of many
coupled oscillators. In an effort to better understand radio tuning,
Adler studied the synchronization of locking circuits, in which a
phase-oscillator synchronizes to a periodic forcing\cite{AdlerProcIRE1946}.
Burykin and Buchman discussed the possibly lethal outcome of the lack
of synchronization among organ systems when taking a patient off of
a mechanical respirator\cite{BurykinComplexity2008}. Gintautus et
al. found synchronization in so-called mixed-reality states, in which
virtual and real systems are coupled and interact in real time\cite{GintautasPhysRevE2007}.
All of these systems exhibit synchronization in some sense. Although
we find these systems to be interesting, the work presented here is
motivated by the many examples listed in the first paragraph: spontaneous
collective behavior of many coupled oscillators in the absence of
external forcing.

In this paper we present yet another system that exhibits synchronization:
small mechanical vibrators coupled through a resonant plate. In addition
to being inexpensive and easy to study, this system provides a unique
twist to the standard coupled-oscillator problem in that the coupling
between the oscillators depends on frequency and exhibits a simple
resonance structure. How does frequency dependent coupling effect
the dynamics of coupled oscillators? Unlike most other globally coupled
oscillator systems, we find history-dependent behavior: an ensemble
of oscillators shows hysteresis in the frequency and degree of synchronization,
and an individual oscillator shows hysteresis in its steady-state
frequency.

\section{Experimental Setup}

\begin{figure}
\begin{centering}
\includegraphics[width=3in]{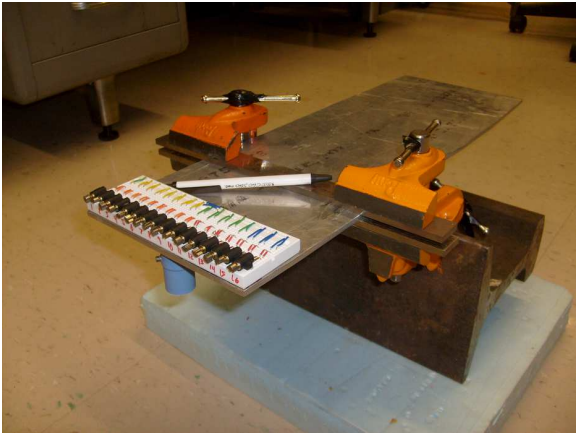}
\par\end{centering}

\begin{centering}
\includegraphics[width=2.5in]{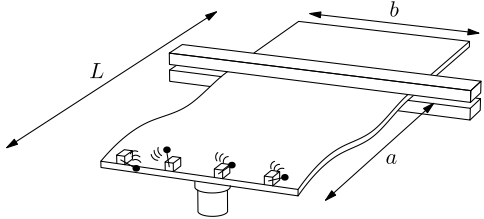}
\par\end{centering}

\caption{\label{fig:Experimental Setup}A photo and diagram of the experimental
setup, as described in the second section.}

\end{figure}

In this work, we study 16 small motors with eccentrically massed rotors.
The motors (All Electronics Corporation, catalog number DCM-204%
\footnote{This item is no longer available in the catalog, but similar motors
can be found in their catalog searching for {}``motor vibrator.''%
}) are small DC motors, the sort used as vibrators in mobile phones.
Each motor has a mass of $3\, g$ and is $2\, cm$ long. Each motor's
rotor has a center of mass that is offset from the axis of rotation,
with a first moment of $0.74\, g\mbox{-}mm$. Vibrations arise from
the rotation of this off-center mass.

To cause the motors to interact, we attach them to a mechanically
compliant and resonant aluminum plate held by clamps as shown in figure~\ref{fig:Experimental Setup}.
The plate is $L=115\, cm$ long, $b=15\, cm$ wide, and $5\, mm$
thick. We adjust the linear response of the system by adjusting the
location where the clamps hold the plate, parametrized by the length
$a$. Although we considered various clamp positions, all of the results
reported here are based on a length of $a=12.5\, cm$, for which the
system has resonances at $68\, Hz$ and $100\, Hz$.

\begin{figure}
\begin{centering}
\includegraphics[width=2.5in]{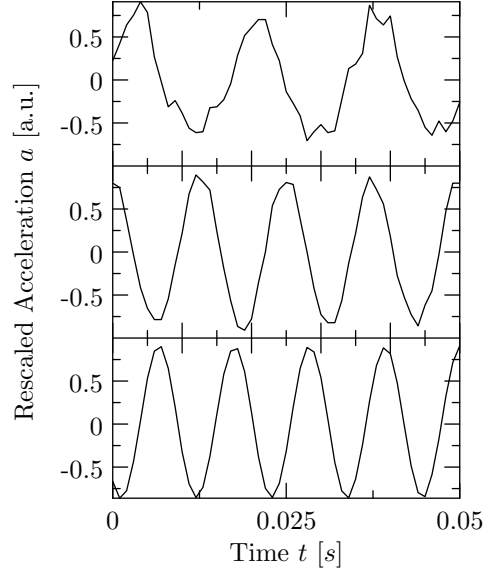}
\par\end{centering}

\caption{\label{fig:Single-motor-data}Typical time series of a single motor
on the plate for different voltages. From top to bottom, the data
correspond to driving voltages are $0.65\, V$, $0.84\, V$, and $1.05\, V$.}

\end{figure}

We measure the plate's vertical acceleration $a\left(t\right)$ using
an accelerometer attached to the plate, a PCB 353B33. In the diagram
shown in figure~\ref{fig:Experimental Setup}, the accelerometer
is depicted by the canister underneath the motors. A few typical time
series of acceleration data due to a single motor are shown in figure~\ref{fig:Single-motor-data}.
The sampling rate for these and all other data we discus is $r=1000\, Hz$.
The plate is a linear medium, so we attribute any observed vibrations
either to the motors or to background sources, such as building vibrations.
We took measurements with a stroboscope to confirm the primary frequencies
obtained from the accelerometer data. In order to reduce spurious
frequencies from the environment, we place the entire setup on a foam
pad. Although some background noise still perturbs the system, these
vibrations do not dominate the signal reported by the accelerometer
and have frequencies much lower than the motors' primary frequencies.

When the motors operate near a resonance, the stroboscope allows us
to observe the plate's mode shape. We find that both resonances, near
$f=68\, Hz$ and $100\, Hz$, do not have any nodes along the array
of motors and that the displacements of all the motors has about equal
magnitude. As such, the coupling between the motors has no appreciable
position dependence.

All of the motors operate from a common power supply but small variations
in each of the motors lead to a distribution of motor frequencies
for a given voltage. We do not attempt to characterize the distribution
of motor speeds in any rigorous way since we are only working with
16 motors. Stroboscopic observations indicate that the frequency distribution
is approximately unimodal.

\section{Behavior of a Single Motor}

\begin{figure}
\begin{centering}
\includegraphics[width=2.5in]{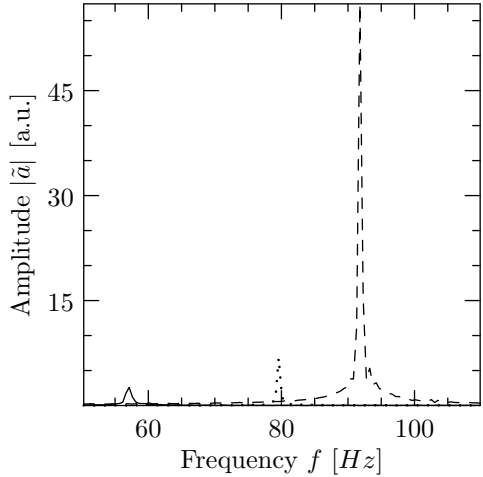}
\par\end{centering}

\caption{\label{fig:Single-Motor-FT}Typical Fourier transforms of a single
motor on the plate for different voltages. The driving voltages are
$0.65\, V$ (--), $0.84\, V$ ($\cdots$), and $1.05\, V$ (- -).}

\end{figure}

\begin{figure}
\begin{centering}
\includegraphics[width=2.5in]{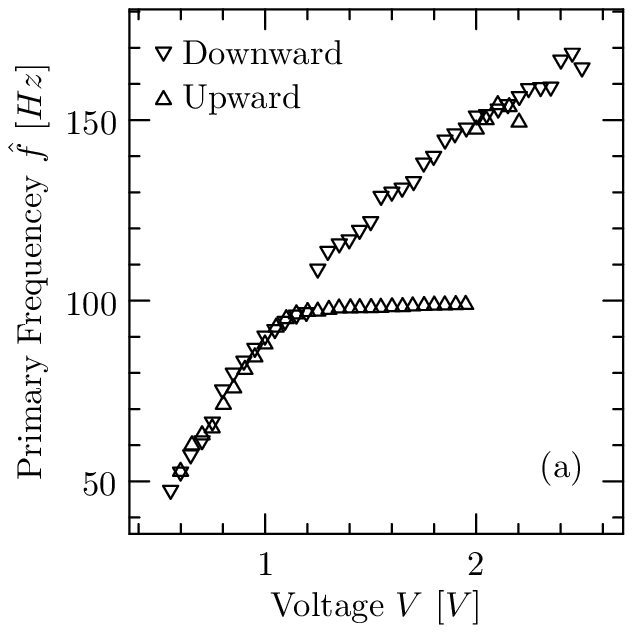} \includegraphics[width=2.5in]{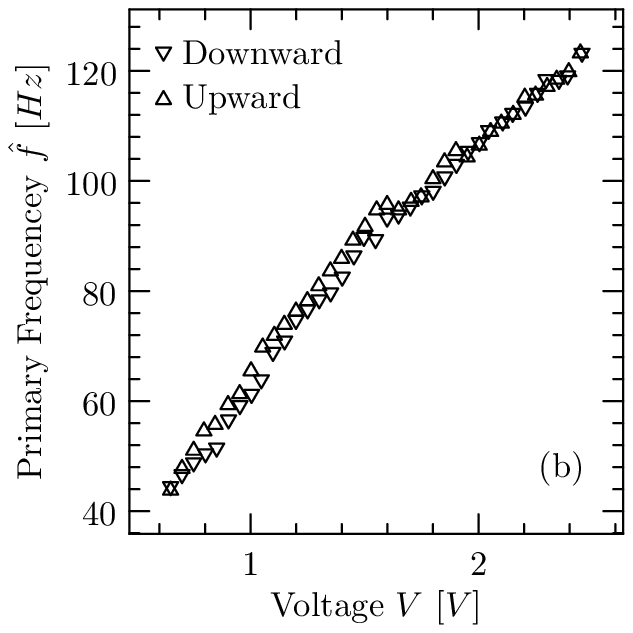}
\par\end{centering}

\caption{\label{fig:Single motor response, Freq vs Voltage}Frequency response
of a single motor verses voltage, both (a) on a resonant plate, and
(b) for comparison, on a rigid support. The motor studied in figure~(a)
was different from the motor studied in figure~(b).}

\end{figure}

In order to discuss how multiple motors interact we must first understand
how a single motor behaves and the sort of response it produces.

\begin{figure}
\begin{centering}
\includegraphics[width=2.5in]{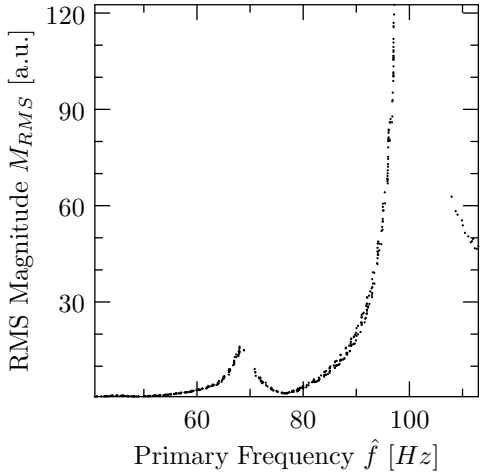}
\par\end{centering}

\caption{\label{fig:Single motor Amplitude vs Freq}Accelerometer amplitude
as a function of motor frequency when being driven by a single motor.
The peaks near 68 $Hz$ and 97 $Hz$ correspond with peaks in the
support's Green function at the same frequencies.}

\end{figure}

Figures~\ref{fig:Single-motor-data} and \ref{fig:Single-Motor-FT}
demonstrate typical%
\footnote{Although the response as a function of primary frequency is typical,
the primary frequency as a function of voltage is not. This is one
of our faster motors.%
} single motor data at voltages $V=0.65\, V$, $0.84\, V$, and $1.05\, V$.
Both figures show stable periodic behavior. The spectra $\tilde{a}\left(f\right)$
in figure~\ref{fig:Single-Motor-FT} were computed from two-second
long data sets $a\left(t\right)$ by short-time Fourier Transform:
\begin{equation}
\tilde{a}\left(f\right)=\int_{T}a\left(t\right)e^{i\,2\,\pi\, f}dt,\end{equation}
as implemented with an FFT. Although the motor's velocity can drift
under special circumstances, the narrow widths of the peaks in figure~\ref{fig:Single-Motor-FT}
demonstrate that a motor's velocity is relatively stable. We identified
the primary frequency $\hat{f}$ of the motor in these and many other
similar experiments by fitting the peak to a Lorentzian. When we plot
the primary frequencies verses the driving voltage we obtain the plot
shown in figure~\ref{fig:Single motor response, Freq vs Voltage}(a),
which we will discuss in more detail below. We also see that the magnitude
of the response varies substantially for different primary frequencies.
Although the magnitude of the signal can be measured by its maximum
value, a more robust measure is the RMS of the Fourier transform in
the vicinity of the peak, $M_{RMS}$. The RMS magnitude can be plotted
against the voltage $V$, but it is better understood as a function
of the primary frequency $\hat{f}$, as shown in figure~\ref{fig:Single motor Amplitude vs Freq}.
We will return to the magnitude measurements shortly, but first we
will examine how the primary frequency depends on the applied voltage.

Although the motor's primary frequency is relatively stable when the
voltage is fixed, figure~\ref{fig:Single motor response, Freq vs Voltage}(a)
shows that the motor's frequency is not a simple function of voltage.
We can compute the primary frequency $\hat{f}$ and the width of the
peak $\delta\hat{f}$ by fitting the magnitude of the spectrum $\tilde{a}$
to a Lorentzian, \begin{equation}
\left|\tilde{a}\left(f\right)\right|^{2}\approx\frac{C}{\left(f-\hat{f}\right)^{2}+\delta\hat{f}^{2}},\label{eq:Lorentzian-Fit}\end{equation}
within the vicinity of of the peak. Shown in figure \ref{fig:Single motor response, Freq vs Voltage}(a)
are the primary frequencies for two different sets of consecutive
measurements, one in which we started at $V=2.4\, V$ and slowly decreased
the voltage to $0.6\, V$ (indicated by triangles pointing downward),
and another in which we started the motors at $V=0.6\, V$ and slowly
increased the voltage to $2.4\, V$ (indicated by the triangles pointing
upward). Although the two measurements demonstrate relatively good
agreement below $V=1\, V$ and above $V=2\, V$, we see a hysteresis
between $V=1$ and $2\, V$. The upward data gets stuck near a resonance
of the plate. In contrast, similar data taken from a separate motor
on a rigid support is shown in figure~\ref{fig:Single motor response, Freq vs Voltage}(b),
and we see that in the absence of resonances, a motor's frequency
is nearly linear in the applied voltage. The marked difference indicates
that the motor interacts strongly with the resonances of the plate,
and these interactions lead to the hysteresis observed in figure~\ref{fig:Single motor response, Freq vs Voltage}(a).

The primary frequency's magnitude $M_{RMS}$ shows a strong dependence
on the primary frequency $\hat{f}$, as shown in figure~\ref{fig:Single motor Amplitude vs Freq}.
We compute the RMS magnitude by\begin{equation}
M_{RMS}=\sqrt{\int_{\hat{f}-\Delta f}^{\hat{f}+\Delta f}\left|\tilde{a}\left(f\right)\right|^{2}df},\end{equation}
where the range $\Delta f$ is a multiple of the width $\delta\hat{f}$
determined by the Lorentzian fit, equation \ref{eq:Lorentzian-Fit}.
The values obtained for $M_{RMS}$ are largely independent of the
choice of $\Delta f$ so log as $\Delta f>\delta f$. As shown in
figure~\ref{fig:Single motor Amplitude vs Freq}, the magnitude of
the plate's response to a single motor is not monotonic in frequency.
We can understand this behavior by noting that the plate has resonances
near $70\, Hz$ and $100\, Hz$, so the plate will have larger accelerations
when driven by a motor near these frequencies than when the motor's
frequency is far from the resonances. This data was obtained by powering
different motors -- one at a time -- at various voltages and taking
two-second data sets for each voltage. Although we could seek a relation
between the RMS magnitude and the applied voltage, figure~\ref{fig:Single motor Amplitude vs Freq}
indicates that the RMS magnitude is a function of primary frequency.
Despite overlaying data from motors at various different locations
on the plate, the magnitude as a function of primary frequency is
remarkably consistent, showing that the coupling between the plate
and the motors for our geometry does not depend substantially on the
motor's position. Apart from the gaps in the data for frequencies
just above the two peaks, the magnitudes in this plot are equivalent
to $f^{4}\, G$, where $G$ represents the passive frequency-dependent
Green function of the system. Calculations not shown here indicate
that the gaps are due to motor-plate interactions that make those
frequencies unstable.

\begin{figure}
\begin{centering}
\includegraphics[width=2.5in]{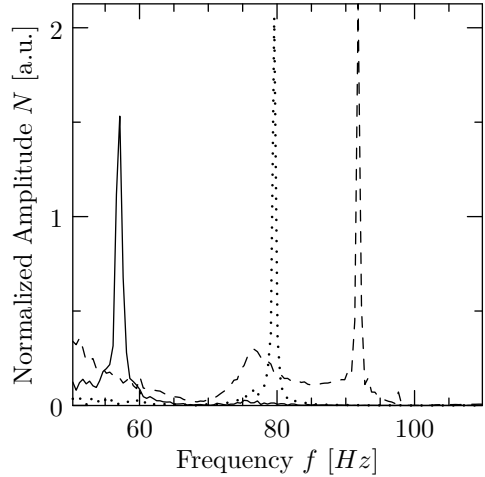}
\par\end{centering}

\caption{\label{fig:Normalize-FT}Normalized plot of data shown in figure~\ref{fig:Single-Motor-FT}.
The driving voltages are $0.65\, V$ (--), $0.84\, V$ ($\cdots$),
and $1.05\, V$ (- -).}

\end{figure}

All of the discussion of data presented so far has focused on single
motors. Since we use a single accelerometer to measure the behavior
of multiple motors acting simultaneously, and since we wish to know
when two motors synchronize, we must obtain a reasonable estimate
for the number of motors at a given frequency. Such an estimate is
not trivial: the resonant response of the plate means that one motor
turning at $95\, Hz$ will produce a much stronger signal than many
synchronized motors with a primary frequency of $78\, Hz$. Our solution
to this problem is to use figure~\ref{fig:Single motor Amplitude vs Freq}
as a normalization curve. We sample the RMS magnitude uniformly --
interpolating where necessary -- to obtain normalization amplitudes
$\hat{M}\left(f\right)$. We then normalize a raw spectrum such as
figure~\ref{fig:Single-Motor-FT} by dividing the amplitudes of the
original spectrum by the normalization amplitudes:\begin{equation}
N\left(f\right)=\frac{\left|\tilde{a}\left(f\right)\right|}{\hat{M}\left(f\right)}.\end{equation}
The result of such a normalization scheme is shown in figure~\ref{fig:Normalize-FT}
for the data presented in figure~\ref{fig:Single-Motor-FT}. Except
for the artifacts at $f=50\, Hz$ and $75\, Hz$ associated with the
signal at $V=1.05\, V$, the scheme appears to work quite well. Even
with the artifacts, single motors can be easily distinguished and
counted, providing us with a decent measure of the number of motors
in the vicinity of a given frequency.

\section{Many Motors on a Resonant Plate}

\begin{figure}
\begin{centering}
\includegraphics[width=3in]{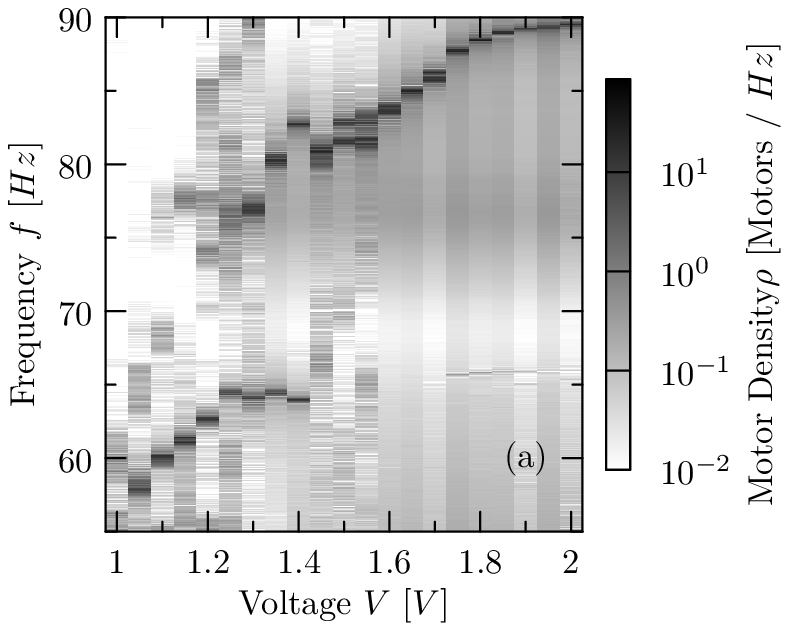} \includegraphics[width=3in]{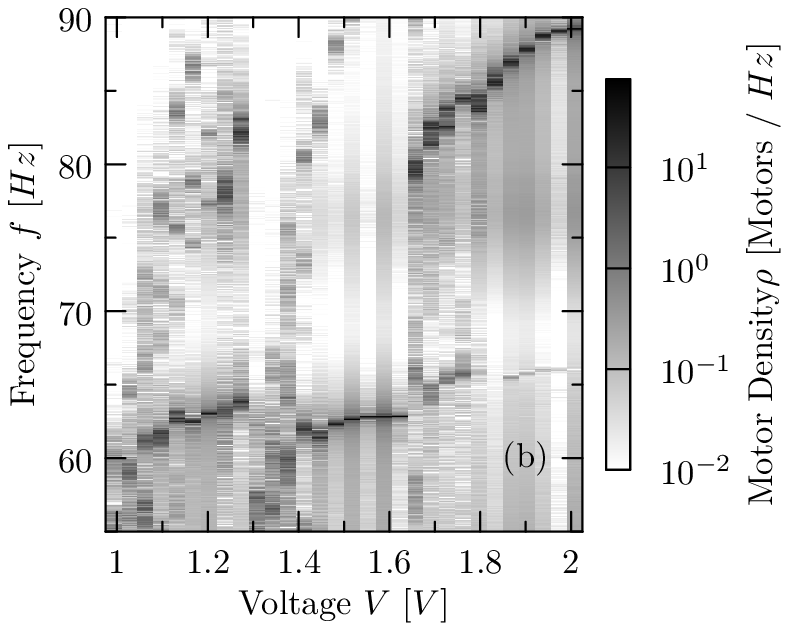}
\par\end{centering}

\caption{\label{fig:Multiple-motors}Behavior of many motors on a plate as
a function of voltage. (a) Behavior as we decrease the voltage starting
from an initially high value. (b) Behavior as we increase the voltage
from an initially low value.}

\end{figure}

\begin{figure}
\includegraphics[width=3in]{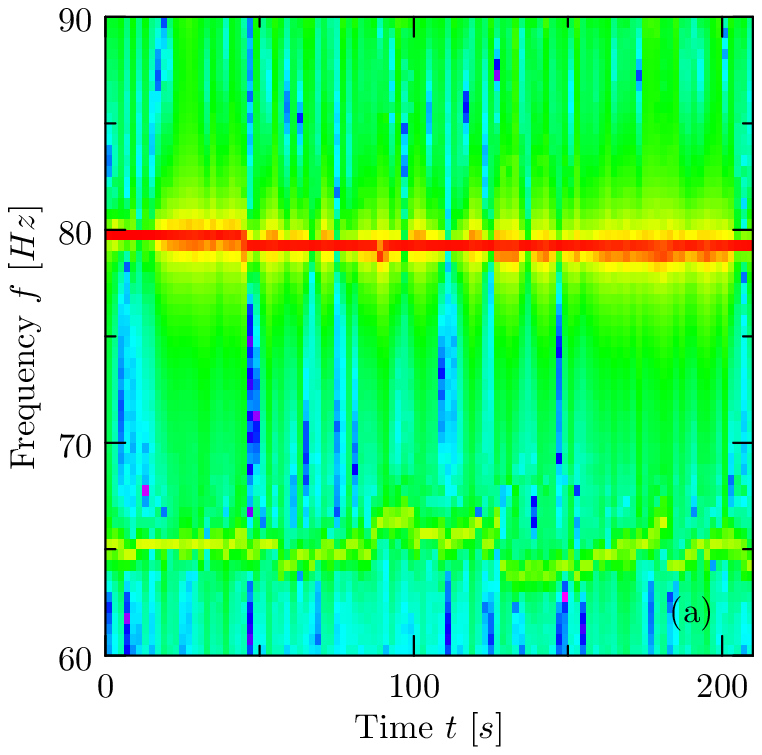} \includegraphics[width=3in]{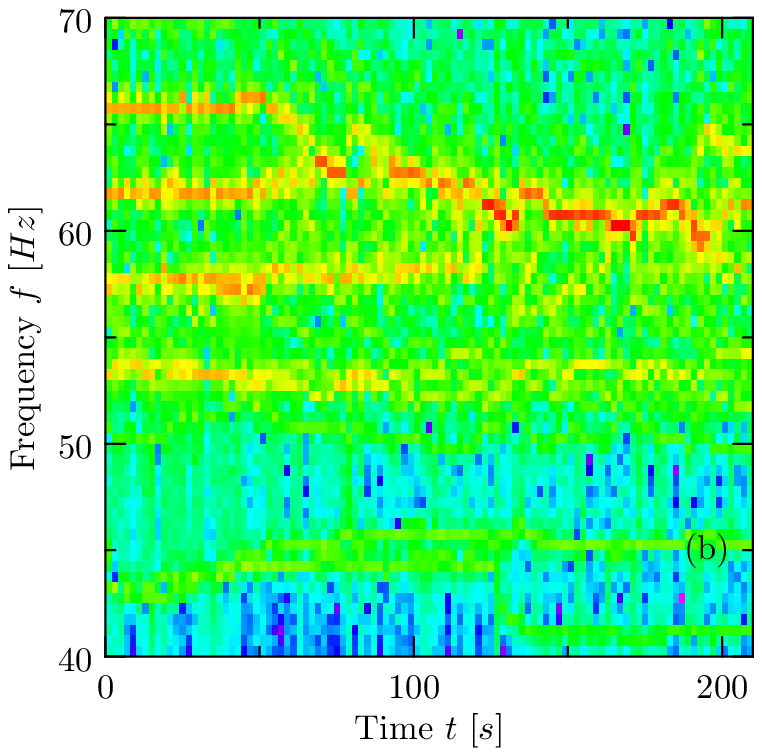}

\caption{\label{fig:Spectrogram-of-multiple}Normalized spectrograms of typical
and atypical dynamics of multiple motors on a resonant plate. (a)
Behavior at $1.49\, V$. (b) Behavior at $1.06\, V$.}

\end{figure}

The essential behavior of multiple motors interacting on the plate
is given in figure~\ref{fig:Multiple-motors}. These plots are consecutive
minute-long measurements that have been Fourier transformed and normalized
as discussed in the previous section. Instead of plotting individual
spectra, like those in figure~\ref{fig:Normalize-FT}, we plot consecutive
spectra by creating gray-scale columns and laying them out sequentially
in order of applied voltage. The difference between figures~\ref{fig:Multiple-motors}(a)
and \ref{fig:Multiple-motors}(b) is that in the former we started
the system at high voltage and stepped the voltage down each consecutive
measurement, whereas in the latter we started the system at low voltage
and stepped the voltage up each consecutive measurement, in a manner
similar to the two data sets shown in figure~\ref{fig:Single motor response, Freq vs Voltage}.
Our motors have multiple stable states for a given voltage, but generally
the motors show partial or full synchronization near $f=65\, Hz$
($1.2\, V<V<1.4\, V$), partial synchronization below $f=85\, Hz$
($1.2\, V<V<1.7\, V$), and nearly full synchronization above $f=85\, Hz$
($V>1.7\, V$). In light of the multiple stable states of a single
motor near a resonance, the multiple stable states of the many-motor
system between $V=1.2\, V$ and $1.7\, V$ is not surprising.

However, the frequency of the resonance that causes the hysteresis
is surprising. Although the magnitude measurements in figure~\ref{fig:Single motor Amplitude vs Freq}
clearly show the resonance near $68\, Hz$, the individual motor's
behavior shown in figure~\ref{fig:Single motor response, Freq vs Voltage}
indicates that the resonance has no noticeable effect on the motor's
frequency. The pronounced effect of the resonance in figures~\ref{fig:Multiple-motors}(a)
and (b) suggest that a resonance's effect on a motor's steady-state
frequency depends on the number of motors near the resonance. We assert
that when we have more motors near a resonance, the resonance's pull
is stronger. To confirm this assertion, note that in figure~\ref{fig:Multiple-motors}(b),
between $V=1\, V$ and $1.1\, V$, there is a motor turning with increasing
frequencies, from $f=65\, Hz$ to $70\, Hz$. At $1.15\, V$ its frequency
is $74\, Hz$. Before proceeding to $1.2\, V$, we forced the motor
back down to the ensemble near $62\, Hz$. Had we continued the measurements
with that motor left unchecked, as we did in other measurements, the
synchronization at $62\, Hz$ would have dispersed at $V=1.45\, V$
or $1.5\, V$. This difference in behavior confirms that the stability
of a resonance depends on the number of motors within a vicinity of
the resonant frequency.

Figure~\ref{fig:Spectrogram-of-multiple} shows the dynamics of the
motors at a fixed voltage over long periods. The stability of the
14 motors synchronized at $f=80\, Hz$ is typical for synchronized
motors, while the behavior of the barely-visible motor wandering near
$65\, Hz$ is typical of unsynchronized motors. The behavior captured
in figure \ref{fig:Spectrogram-of-multiple}(b) is unusually dynamic
for \emph{steady-state} behavior: it demonstrates nearly all of the
typical \emph{transient} behavior that the motors exhibit at higher
voltages. The most interesting behavior is the synchronization and
desynchronization of motors spinning faster than $f=55\, Hz$. In
particular, two groups of motors gradually merge between $t=50\, s$
and $70\, s$, and then merge with yet another group to form a very
large group at $130\, s$. This group is not stable and some of the
motors split off just before $200\, s$ before re-merging about a
minute later (not shown).

\section{Conclusion and Future Work}

We set out to answer this question: How does frequency-dependent coupling
effect the synchronization dynamics of many coupled oscillators? We
found that the behavior of individual motors and ensembles of motors
interacting with a resonant plate is reproducible but history dependent.
However, much remains to be known.

In the work presented here, we studied the behavior of one motor on
a rigid support and others on a resonant plate. In future experimental
work, one could measure the same motor's angular velocity when it
is on a rigid support and on a plate to get a direct comparison of
frequency as a function of voltage. We also considered a specific
geometry for the plate and motors so that all of the motors interacted
with the plate in nearly the same way. How would the motors behave
differently if some of them were placed on the nodes for a given resonance?

We have not presented a model for our system, but many of the behaviors
exhibited in this paper seem theoretically tractable. These include
the stability criteria for a single motor as it approaches a resonance
from above or settles on a resonance from below, the stability criteria
for synchronization near a resonance, and the steady-state dynamics
of many motors at weak coupling. Is the behavior at weak coupling
chaotic? Is it stochastic? What governs the time scales of merging
and collapsing groups?

We look forward to studying all of these fascinating behaviors.

\subsubsection*{Acknowledgments}

We are thankful to John Kolinski for initiating the motors project
and Nick Wolff for useful discussion. This work was supported in part
by NSF grant 0528096.

		\bibliographystyle{ieeetr}
\bibliography{Sources}

\end{document}